\documentclass[%
 reprint,
 amsmath,amssymb,
 aps,
]{revtex4-2}

\usepackage[export]{adjustbox}
\usepackage{relsize}
\usepackage{graphicx}
\usepackage{bm}
\usepackage{graphicx}
\usepackage{comment}
\usepackage[caption=false]{subfig}
\usepackage{rotating}
\usepackage[]{hyperref}
\hypersetup{
colorlinks,
linkcolor={blue},
linktocpage=true,
citecolor={red},
urlcolor={green}
}

\newcommand{\MuToEmConv}{$\mu^-N \rightarrow e^- N$}
\begin{document}

\title{The Advanced Muon Facility: a proposed multi-purpose muon facility at Fermilab}
\author{Sophie Middleton}
\date{\today}

\begin{abstract}
\textbf{Contribution to the 25th International Workshop on Neutrinos from Accelerators}\\
Charged lepton flavor violation (CLFV) is expected in a diverse set of new physics scenarios. The current generation of experiments probe CLFV in the muon sector in three complementary channels: $\mu^-N \rightarrow e^- N$ (Mu2e, COMET), $\mu^+ \rightarrow e^+ \gamma$ (MEG-II), and $\mu^+ \rightarrow e^+e^+e^-$s (Mu3e). These experiments aim to enhance existing limits by several orders-of-magnitude in the coming decade and offer discovery potential to many new physics models. The proposed Advanced Muon Facility (AMF) would be a multi-purpose muon facility based at Fermilab and introduces an innovative approach based on a muon storage ring to enable a full suite of muon CLFV experiments. AMF would host CLFV experiments with sensitivities orders-of-magnitude beyond the present era. In the event of a signal in these currently planned experiments, AMF would enable additional measurements to elucidate the nature of the new physics observed. The design and R$\&$D for AMF is in its infancy. This article outlines the motivations for AMF, detailing on-going R$\&$D efforts, and highlighting potential synergies with the proposed muon collider.
\end{abstract}

\maketitle

\section{Motivations}

Charged lepton flavor violation (CLFV) has not yet been observed. The introduction of non-zero neutrino masses and mixing angles provides a mechanism for CLFV in the Standard Model (SM) through mixing in loops. However, if neutrino masses arise from Yukawa interactions with the Higgs boson, these SM CLFV rates are GIM-suppressed, resulting in branching fractions of the order of $\mathcal{O}(10^{-54})$ \cite{Marciano2008}, far beyond the reach of any conceivable experiment. Therefore, any experimental observation of a CLFV process would constitute a discovery of beyond Standard Model (BSM) physics. 

Muons have consistently provided powerful constraints on CLFV \cite{bernstein_history}. Muons do not decay hadronically, resulting in much simpler final states than those of the heavier tau lepton. In addition,  they have a relatively long lifetime of 2.2 $\mu$s and intense muon beams can be produced as by-products from high energy protons on target at facilities such as PSI, J-PARC and Fermilab, providing the high statistics required to investigate processes with potentially very small rates.  There are three primary channels used to search for CLFV in the form of $\mu \rightarrow e$:

\begin{itemize}
    \item \textbf{$\mu^+ \rightarrow e^+ \gamma$}: The MEG-II experiment at PSI has recently published an updated combined (with MEG) limit: BR($\mu^+ \rightarrow e \gamma) < 3.1 \times 10^{-13}$ at 90\% C.L.\cite{MEGII:2023ltw}. MEG-II aims to acquire enough data to provide an upper limit of $< 6 \times 10^{-14}$ at 90 $\%$ C.L by the end of its run period in the late 2020s \cite{MEGII:2021fah}.
    \item \textbf{$\mu^+ \rightarrow e^+e^+e^-$}: This process complements $\mu \rightarrow e \gamma$, in processes dominated by $\gamma$-penguin diagrams .e.g. SUSY loops, the $\mu \rightarrow e \gamma$ rate is larger, however, $\mu \rightarrow e \gamma$ is insensitive to $Z$ or $H$ penguins or leptoquark processes. The current limit is BR($\mu \rightarrow eee < 1.0 \times 10^{-12}$ ) at 90\% C.L. \cite{BELLGARDT19881} and the Mu3e experiment, also at PSI, will improve the limit by up to four orders-of-magnitude \cite{hesketh2022mu3e}. Mu3e plans to begin data taking in the mid-2020s.
     \item $\mu^-N \rightarrow e^- N$: The $\mu^-$ is captured by an atomic nucleus and cascades into a muonic 1s state.  A coherent interaction with the nucleus results in muon-to-electron conversion with an outgoing monoenergetic electron with energy at just less than the muon mass. The coherent process is enhanced by $Z^2$ with respect to the incoherent process. Both Mu2e (at Fermilab) and COMET (at J-PARC) will search for the conversion process using an aluminum nucleus, both aiming for a sensitivity of four orders-of-magnitude below the present upper limit ($7 \times 10^{-13}$ at 90 $\%$ C.L from SINDRUM-II\cite{Bertl2006}). Mu2e and COMET will begin data-taking in the mid-2020s and run for several years to acheive their goals.
\end{itemize}

Experiments searching for the first two decay channels use quasi-continuous, low-momentum, $\mu^{+}$ beams while those searching for the third process utilize an intense, pulsed, $\mu^{-}$ beam.  The choice of beam is to aid the physics, the use of a quasi-continuous beam in decay experiments suppresses accidental backgrounds, while in conversion searches the pulsed beam is used to avoid prompt beam backgrounds overwhelming the signal. The decay experiments use a $\mu^{+}$ beam to avoid muon capture by the target nuclei which distorts the decay kinematics. Many well-motivated extensions of the SM predict much higher rates of the CLFV processes than those predicted in the minimally extended SM, many suggest rates that fall within the projected reach of these upcoming experiments. Examples include: models of SO(10) supersymmetry \cite{Calibbi2014,Calibbi2006}, scalar lepto-quarks \cite{Arnold, heeck} and those with additional Higgs doublets \cite{Abe2017}. Searches for \MuToEmConv~ can also help place limits on lepton flavor violating Higgs decays, Ref.~\cite{harnik} suggests that Mu2e and COMET have sensitivity up to $BR (h \rightarrow \mu e)$ of $10^{-10}$. CLFV measurements can also help elucidate the mechanism behind neutrino masses, as different mass generating Lagrangians predict different rates of CLFV, therefore, determining the rates of multiple CLFV processes allows us to discriminate models~\cite{HAMBYE201413}. 

In order to completely explore all possible new physics scenarios it is necessary to look for all three CLFV processes (as well as the analogous processes involving taus and LFV Higgs decays). New physics contributions to CLFV in dimension-6 operators can be divided into dipole, or photonic contributions, where the CLFV is mediated by a photon that is absorbed by the nucleus, and contact interactions, where a new heavy virtual particle is exchanged which couples to the fermions. Searches for radiative $\mu^{+} \rightarrow e^{+}  \gamma$ decays are only sensitive to electromagnetic dipole interactions. On the other hand, experiments searching for \MuToEmConv~ and $\mu^{+} \rightarrow e^{+} e^{-} e^{+}$ are sensitive to processes that result from either photonic or four fermion (contact) type interactions.  While searches for $\mu^{+} \rightarrow e^{+} \gamma$ provide powerful constraints on the dipole operators, \MuToEmConv~ conversion is the most sensitive observable to explore operators involving quarks. Reference~\cite{Davidson:2022nnl} details how these upcoming experiments provide a sensitivity to effective new physics masses up to $\mathcal{O}(10^{4})$ TeV/c$^{2}$, far beyond what is achievable at direct searches at colliders. The relative contributions to the different CLFV channels reflects the nature of the underlying physics model responsible \cite{calibbi2017charged}. In the event that CLFV signals are observed in the upcoming experiment, comparisons among the results from experiments dedicated to different CLFV channels can help elucidate the nature of new physics responsible. In addition, the rate of \MuToEmConv~ is dependent on the atomic nucleus   ~\cite{kitano2002,Cirigliano2009,NucleusDependence,Leo}. A way to elucidate a conversion signal at Mu2e and/or COMET would be to measure the conversion rate in a material complementary to aluminum. 

\section{The Advanced Muon Facility}

The Advanced Muon Facility (AMF)  \cite{CGroup:2022tli} is an ambitious proposed high-intensity muon science complex. AMF would deliver the world's most intense $\mu^{+}$ and $\mu^{-}$ beams, enabling a broad muon science program. AMF would host a flagship CLFV program, searching for all three muon CLFV channels, improving upon the sensitivity of planned experiments by orders-of-magnitude or further constraining the nature of new physics in the case of an observation. The main physics goals of AMF are:

\begin{itemize}
\item to study \MuToEmConv~ on gold, reaching the highest mass scale effective mass scale of nearly $10^5$ TeV/c$^2$ \cite{Davidson:2022nnl};
\item to achieve limits ${\mathcal O}(10^{-(15--16)})$  in $\mu^{+} \rightarrow e^{+} e^{-} e^{+}$, to match the power of the muon-to-electron conversion search.
\item to make subsequent, higher-statistics, measurements to understand the mechanism responsible in the event that these processes have been observed already in any of the upcoming/current experiments. 
\end{itemize}
AMF would be based at Fermilab, utilizing the PIP-II proton beam and associated accelerator infrastructure. R\&D for AMF has been endorsed by the 2023 P5 \cite{p5}. In terms of the experimental designs, for all three channels, significant changes on the current designs are required to achieve these physics goals.  AMF could also host a muonium-antimuonium oscillation search competitive with MACE \cite{mace} (factor of 100 improvement on current limit), muon collider R$\&$D efforts and other proposed experiments which utilize an intense  $\mu^{+}$ or $\mu^{-}$ beam. 

 In the event of signals in the current era of CLFV experiments, AMF would elucidate these signals by making complementary measurements. It is found that a heavy nucleus such as lead or gold would provide the best comparison to aluminum. Both COMET and Mu2e utilize a pulsed proton beam. This is beneficial for these experiments as it allows them to effectively ``wait out" pion decay through use of a ``delayed live-gate," essentially ignoring anything in their detectors before $\sim$ a few hundred ns after the proton pulse hits the target, at which time all prompt pions have decayed. Taking Mu2e as an example, here the proton pulses are $\sim$ 250 ns wide and separated by 1695 ns. The mean muonic atom lifetime in aluminum is 864 ns, in gold it is 73 ns \cite{suzuki}. With the Mu2e beam, it would be impossible to make an accurate measure of muon conversion in gold, it would be overwhelmed by beam backgrounds. To measure conversion in gold a completely new conceptual design is required.

 AMF follows a similar concept as the proposed PRISM (Phase Rotated Source of IntenSe Muons) \cite{prism}. A high intensity beam (up to 1MW) is incident on a target in a capture solenoid, producing pions. These pions decay to muons and travel through the beamline. They have a broad mommentum and a short time profile. The muons are injected into a Fixed Field Alternating (FFA) gradient machine and undergo a process of ``phase rotation." The beam becomes wide-spread in time, with a small momentum spread. The momentum of the outgoing beam can be controlled, ultimately providing a mono-energetic muon beam at extraction. 

The use of the FFA provides a pure muon beam, without pion contamination; consequently, a ``delayed live-gate" is no longer necessary and conversion in a heavy target such as gold could be observed. Additionally, through using the FFA, we have control over the eventual momentum of the muons. A choice of low central momentum (e.g. 20 -- 40 MeV/c) means thinner targets can be used, reducing multiple scattering and energy losses, and improving the momentum resolution on the outgoing electrons. This is essential to remove decay backgrounds. The PSI decay experiments already utilize a $\sim$ 30 MeV/c surface muon beam.

\begin{figure*}[ht]
  \centering
  \includegraphics[width=0.5\textwidth]{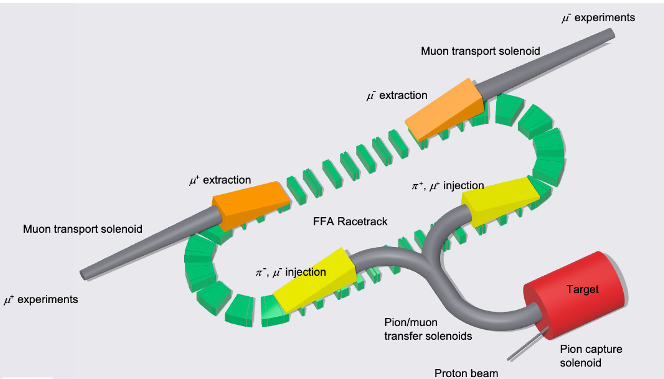}
  \caption{One possible configuration of the AMF facility (not final, illustration only) \cite{david}.}  
  \label{fig:concept}
\end{figure*}

Figure~\ref{fig:concept} shows a possible AMF schematic; this is purely conceptual. A large amount of effort is required to design a feasible lattice for the FFA and technology capable of injection and extraction into the FFA ring.  For AMF, utilizing a race-track FFA \cite{Pasternak:2010zz}, might prove advantageous. The straight sections could allow for easier injection and extraction. Unlike PRISM, which would host only the PRIME conversion search, AMF will host all three muon CLFV searches. As a result, AMF requires both a $\mu^{+}$ and $\mu^{-}$ beam, both with central momenta of 20 -- 40 MeV/c, which is optimal for both the decay and conversion experiments. The two beams could be produced simultaneously, the positive muons circulating one way, and negative muons the other, with the output muon beams sent to different experiments. Alternatively, sequential runs could take place.  This is an ambitious goal, and requires detailed, focused, design, and the input from experts. It would be a completely novel facility. 
\section{Facility Design}
\subsection{Proton Beam}

PIP-II is an upgrade to the Fermilab accelerator complex to increase the beam power available for the high energy neutrino program. The goal is to construct a 800 MeV superconducting linac that will inject protons into the existing 8 GeV Booster, replacing the current 400 MeV linac. The increased injection energy, coupled with a sophisticated transverse painting scheme, will result in a higher-intensity beam. The Booster repetition rate will also be increased from 15 to 20 Hz. The neutrino program will use less than 1$\%$ of PIP-II's capacity. There will potentially be 1.6 MW of 800 MeV beam available for other research opportunities. Hence, we propose AMF be based at Fermilab utilizing this surplus beam.

Driving the proposed  FFA will require a ``compressor ring" to accumulate protons and extract them with the required bunch structure. The FFA will need proton bunches that are on the order of 10 ns long at 100-1000 Hz. The direct output of the PIP-II Linac would be minuscule with that structure. Proposed compressor designs are discussed in detail in ~\cite{CGroup:2022tli, compressor:LOI}.

\subsection{Production Target}

AMF will have an incoming proton beam with 100 kW - 1MW beam power, at a 100-1000 Hz pulse rate. An R$\&$D effort for a 100 kW scale target operating with 800 MeV beam from PIP-II is already underway for the proposed Mu2e upgrade, Mu2e-II \cite{Mu2eII}. The driving constraints on the eventual design will be peak stresses, radiation damage, and heat management. Scaling the target system to $\mathcal{O}$(1MW) will require significant R$\&$D. We can draw on existing experience from neutrino production facilities and spallation sources (e.g. NOvA, LBNF, and SNS) that utilise/will utilise megawatt-scale targets. There are distinct challenges in the AMF case, since the target will be embedded within the small, open bore of a super-conducting solenoid. The solenoid must be protected from the high levels or heat and radiation. There are obvious synergies with target requirements for a proposed muon collider. Muon collider design and prototyping efforts have attempted to address this problem with various systems including liquid metal or granular jets, for example \cite{MCtarget}. 

\subsection{FFA}

FFAs can provide lattices with very large acceptances in both the transverse and longitudinal planes, which is essential for muons, produced as tertiary beams, with large emittances and momentum spread. The AMF muons originate from pions produced when a compressed proton bunch interacts with the production target. These muons are then injected into the FFA that is equipped with an RF system. Phase-space rotation is performed to transform the initial short muon bunch with a large momentum spread of $\sim$ 20 $\%$ $p/p_0$  into a long bunch with the momentum spread reduced by about an order-of-magnitude. The narrow momentum spread beam is then extracted and sent to the experiments. Several lattice solutions have been proposed for PRISM \cite{Alekou:2013eta}, with an initial baseline based on a scaling DFD triplet \cite{Witte:2012zza}, which was successfully constructed and verified experimentally \cite{Sato:2008zze}. The studied solutions included both scaling and non-scaling FFA lattices, consisting of regular cells, and a racetrack geometry. The racetrack geometry has the advantage of long straight sections, seemingly making injection and extraction easier. Significant effort is required in collaboration with FFA experts to design the lattice for AMF. A series of workshops and discussions, bringing together global experts is expected in the coming years.

A conceptual design for the beam transport and injection system in to the FFA has been proposed in \cite{CGroup:2022tli}; feasibility studies must follow. In theory, the extraction system could be a reverse copy of this proposed injection system, but may be simpler as the momentum spread much smaller. The main challenge is the rise time of the extraction kicker(s), which needs to be shorter than the injection ones as the bunch at extraction is significantly longer. Studies must also be done to understand the optimal way of circulating beams of opposite charges simultaneously. 

\section{Experiment Designs}

The suite of individual experiments also require significant re-design to fully utilize AMF to achieve their proposed physics goals. 

\subsection{Decay Experiments}

The current decay experiments exploit surface muons produced by the decay of pions at rest on the surface of a production target. The resulting 29.8 MeV/c muons are further slowed down with a degrader to improve the stopping efficiency in a very thin target. For $\mu^{+} \rightarrow e^{+} \gamma$,  to reach a sensitivity $< 10^{-14}$ requires a calorimeter with a much larger acceptance than the MEG-II LXe detector. An evolution could use thin layers of high-Z material to convert photons into $e^{+}e^{-}$ pairs, which could be accurately reconstructed to obtain a precise measurement of the photon energy and the position of the conversion point. Additionally, the MEG-II drift chamber is expected to experience a gain loss of 25$\%$ per year the beam rates of $7 \times 10^{7} \mu$/s. There are several proposed solutions, all need further study. Silicon pixel trackers could be a solution but the material budget would need to be kept under control. 50 $\mu$m thick HV-MAPS pixel sensor have recently been developed for the Mu3e experiment, these could be adapted, but it would be beneficial to reduce the thickness to 25 $\mu$m. \cite{CGroup:2022tli} A significant R$\&$D program is required to fully understand the performance of these concepts and develop a complete experimental design capable of operating in the intense AMF environment.

\subsection{Conversion Experiment}

At AMF the impact of beam-induced backgrounds would be vastly reduced due to the long muon storage time. Therefore, an AMF-based conversion experiment could use high-Z target material, such as gold. The increase in proton beam power, in conjunction with the muon cooling capabilities of the FFA, will also greatly increase the number of stopped muons in the target, providing the opportunity to probe rates down to $10^{-19}$. 

The key to accurately measuring a conversion signal comes from discriminating the monoenergetic conversion signal from the SM decay-in-orbit (DIO) electrons from the target. The momentum spectrum associated with DIO electrons has a long recoil tail that can come close to the conversion signal. Distinguishing these background electrons from the signal electrons, necessitates very good momentum resolution in the tracking detectors ($<$ 180 keV/c at Mu2e for example). Through utilising a low central momentum (20 - 40 MeV/c) a thin target can be used, minimizing straggling and energy losses of the outgoing electrons at the target, this is a significant factor in reducing the resolution at existing experiments. Cosmic rays are also a significant source of background events, but recent studies of a proposed Mu2e-II experiment have shown that this level of performance is within reach \cite{Mu2eII}.  There are several candidates for high-performance, low-mass trackers, including the proposed Mu2e-II straw-tracker with 8 $\mu$m wall thickness \cite{Mu2eII}, low-mass silicon sensors, such as HV-MAPS or micro-pattern gas detectors proposed for the Belle-II tracking TPC, and even exotic novel material \cite{hoeferkamp2022novelsensorsparticletracking}. With regards to the other two major detector systems, the calorimeter and cosmic ray active veto, potential technologies are being investigated by the Mu2e-II effort \cite{Mu2eII}. The development of a detector concept to take full advantage of AMF should be a manageable challenge, but requires significant R$\&$D.

\section{Conclusions}

To summarize, AMF is a novel facility that would provide world-leading science, utilizing an intense muon beam. The current concept envisions both $\mu^{+}$ and $\mu^{-}$ beams and the primary goal is to host all three muon CLFV searches. There is clear motivation to build such a facility, no matter what the outcome of the current era of CLFV experiments. There is a significant amount of R$\&$D required to complete the conceptual design of AMF and the associated experiments. R$\&$D for AMF was endorsed by the 2023 P5 report \cite{p5}. Synergies with the muon collider R$\&$D can be exploited to engage experts, the eventual goal is to provide a conceptual design of AMF by the end of this decade. 

\bibliographystyle{utcap_mod}
\bibliography{paper}

\providecommand{\href}[2]{#2}\begingroup\raggedright\begin{thebibliography}{10}

\bibitem{Marciano2008}
W.~J. Marciano, T.~Mori, and J.~M. Roney \href{http://dx.doi.org/10.1146/annurev.nucl.58.110707.171126}{Ann. Rev. Nucl. Part. Sci. {\bfseries 58} (2008) 315--341}.

\bibitem{bernstein_history}
R.~H. Bernstein \href{http://dx.doi.org/10.3389/fphy.2019.00001}{Frontiers in Physics {\bfseries 7} (2019) 1}. \url{https://www.frontiersin.org/article/10.3389/fphy.2019.00001}.

\bibitem{MEGII:2023ltw}
{\bfseries MEG II} Collaboration, K.~Afanaciev {\em et~al.} \href{http://dx.doi.org/10.1140/epjc/s10052-024-12416-2}{Eur. Phys. J. C {\bfseries 84} no.~3, (2024) 216}, \href{http://arxiv.org/abs/2310.12614}{[{\ttfamily 2310.12614}]}. [Erratum: Eur.Phys.J.C 84, 1042 (2024)].

\bibitem{MEGII:2021fah}
{\bfseries MEG II} Collaboration, A.~M. Baldini {\em et~al.} \href{http://dx.doi.org/10.3390/sym13091591}{Symmetry {\bfseries 13} no.~9, (2021) 1591}, \href{http://arxiv.org/abs/2107.10767}{[{\ttfamily 2107.10767}]}.

\bibitem{BELLGARDT19881}
U.~Bellgardt {\em et~al.} \href{http://dx.doi.org/https://doi.org/10.1016/0550-3213(88)90462-2}{Nuclear Physics B {\bfseries 299} no.~1, (1988) 1--6}. \url{https://www.sciencedirect.com/science/article/pii/0550321388904622}.

\bibitem{hesketh2022mu3e}
{\bfseries Mu3e} Collaboration, G.~Hesketh, S.~Hughes, A.-K. Perrevoort, and N.~Rompotis, ``{\em {The Mu3e Experiment}},'' in {\em {Snowmass 2021}}.
\newblock 4, 2022.
\newblock \href{http://arxiv.org/abs/2204.00001}{[{\ttfamily 2204.00001}]}.

\bibitem{Bertl2006}
{\bfseries SINDRUM II} Collaboration, W.~H. Bertl {\em et~al.}
\href{http://dx.doi.org/10.1140/epjc/s2006-02582-x}{Eur. Phys. J. {\bfseries C47} (2006) 337--346}.

\bibitem{Calibbi2014}
L.~Calibbi, P.~Paradisi, and R.~Ziegler \href{http://dx.doi.org/10.1140/epjc/s10052-014-3211-x}{Eur. Phys. J. C {\bfseries 74} no.~12, (Dec, 2014) }. \url{http://dx.doi.org/10.1140/epjc/s10052-014-3211-x}.

\bibitem{Calibbi2006}
L.~Calibbi, A.~Faccia, A.~Masiero, and S.~K. Vempati \href{http://dx.doi.org/10.1103/PhysRevD.74.116002}{Phys. Rev. D {\bfseries 74} (Dec, 2006) 116002}. \url{https://link.aps.org/doi/10.1103/PhysRevD.74.116002}.

\bibitem{Arnold}
J.~M. Arnold, B.~Fornal, and M.~B. Wise \href{http://dx.doi.org/10.1103/PhysRevD.88.035009}{Phys. Rev. D {\bfseries 88} (2013) 035009}, \href{http://arxiv.org/abs/1304.6119}{[{\ttfamily 1304.6119}]}.

\bibitem{heeck}
J.~Heeck and D.~Teresi \href{http://dx.doi.org/10.1007/JHEP12(2018)103}{JHEP {\bfseries 12} (2018) 103}, \href{http://arxiv.org/abs/1808.07492}{[{\ttfamily 1808.07492}]}.

\bibitem{Abe2017}
T.~Abe, R.~Sato, and K.~Yagyu \href{http://dx.doi.org/10.1007/JHEP07(2017)012}{JHEP {\bfseries 07} (2017) 012}, \href{http://arxiv.org/abs/1705.01469}{[{\ttfamily 1705.01469}]}.

\bibitem{harnik}
R.~Harnik, J.~Kopp, and J.~Zupan \href{http://dx.doi.org/10.1007/JHEP03(2013)026}{JHEP {\bfseries 03} (2013) 026}, \href{http://arxiv.org/abs/1209.1397}{[{\ttfamily 1209.1397}]}.

\bibitem{HAMBYE201413}
T.~Hambye \href{http://dx.doi.org/10.1016/j.nuclphysbps.2014.02.004}{Nucl. Phys. B Proc. Suppl. {\bfseries 248-250} (2014) 13--19}, \href{http://arxiv.org/abs/1312.5214}{[{\ttfamily 1312.5214}]}.

\bibitem{Davidson:2022nnl}
S.~Davidson and B.~Echenard \href{http://dx.doi.org/10.1140/epjc/s10052-022-10773-4}{Eur. Phys. J. C {\bfseries 82} no.~9, (2022) 836}, \href{http://arxiv.org/abs/2204.00564}{[{\ttfamily 2204.00564}]}.

\bibitem{calibbi2017charged}
L.~Calibbi and G.~Signorelli \href{http://dx.doi.org/10.1393/ncr/i2018-10144-0}{Riv. Nuovo Cim. {\bfseries 41} no.~2, (2018) 71--174}, \href{http://arxiv.org/abs/1709.00294}{[{\ttfamily 1709.00294}]}.

\bibitem{kitano2002}
R.~Kitano, M.~Koike, and Y.~Okada \href{http://dx.doi.org/10.1103/PhysRevD.66.096002}{Phys. Rev. D {\bfseries 66} (Nov, 2002) 096002}. \url{https://link.aps.org/doi/10.1103/PhysRevD.66.096002}.

\bibitem{Cirigliano2009}
V.~Cirigliano, R.~Kitano, Y.~Okada, and P.~Tuzon \href{http://dx.doi.org/10.1103/PhysRevD.80.013002}{Phys. Rev. D {\bfseries 80} (2009) 013002}, \href{http://arxiv.org/abs/0904.0957}{[{\ttfamily 0904.0957}]}.

\bibitem{NucleusDependence}
S.~Davidson, Y.~Kuno, and M.~Yamanaka \href{http://dx.doi.org/10.1016/j.physletb.2019.01.042}{Phys. Lett. B {\bfseries 790} (2019) 380--388}, \href{http://arxiv.org/abs/1810.01884}{[{\ttfamily 1810.01884}]}.

\bibitem{Leo}
L.~Borrel, D.~G. Hitlin, and S.~Middleton \href{http://arxiv.org/abs/2401.15025}{[{\ttfamily 2401.15025}]}.

\bibitem{CGroup:2022tli}
M.~Aoki {\em et~al.} in {\em {Snowmass 2021}}.
\newblock 3, 2022.
\newblock \href{http://arxiv.org/abs/2203.08278}{[{\ttfamily 2203.08278}]}.

\bibitem{p5}
P5, ``{\em 2023 P5 Report:Pathways to Innovation and Discovery in Particle Physics Report of the 2023 Particle Physics Project Prioritization Panel},'' 2023.
\newblock \url{https://www.usparticlephysics.org/2023-p5-report/}.

\bibitem{mace}
A.-Y. Bai {\em et~al.}, ``{\em {Snowmass2021 Whitepaper: Muonium to antimuonium conversion}},'' in {\em {Snowmass 2021}}.
\newblock 3, 2022.
\newblock \href{http://arxiv.org/abs/2203.11406}{[{\ttfamily 2203.11406}]}.

\bibitem{suzuki}
T.~Suzuki, D.~F. Measday, and J.~P. Roalsvig \href{http://dx.doi.org/10.1103/PhysRevC.35.2212}{Phys. Rev. C {\bfseries 35} (Jun, 1987) 2212--2224}. \url{https://link.aps.org/doi/10.1103/PhysRevC.35.2212}.

\bibitem{prism}
Y.~Kuno, ``{\em {PRISM/PRIME}},'' \href{http://dx.doi.org/10.1016/j.nuclphysbps.2005.05.073}{Nucl. Phys. B Proc. Suppl. {\bfseries 149} (2005) 376--378}.

\bibitem{Pasternak:2010zz}
J.~Pasternak {\em et~al.} Conf. Proc. C {\bfseries 100523} (2010) WEPE056.

\bibitem{david}
D.~Hitlin, ``{\em The Advanced Muon Facility},'' 2023.
\newblock \url{https://indico.desy.de/event/37920/timetable/?view=standard}.

\bibitem{compressor:LOI}
E.~Prebys, R.~Bernstein, and J.~Pasternak, ``{\em {Letter of Interest: Bunch Compressor for the PIP-II Linac}},'' 2021.
\newblock \url{{https://www.snowmass21.org/docs/files/summaries/AF/SNOWMASS21-AF5_AF0-RF5_RF0_Prebys-071.pdf}}.

\bibitem{Mu2eII}
{\bfseries Mu2e-II} Collaboration, K.~Byrum {\em et~al.}, ``{\em {Mu2e-II: Muon to electron conversion with PIP-II}},'' in {\em {Snowmass 2021}}.
\newblock 3, 2022.
\newblock \href{http://arxiv.org/abs/2203.07569}{[{\ttfamily 2203.07569}]}.

\bibitem{MCtarget}
M.~Calviani {\em et~al.}, ``{\em Target Development for Muon Collider},'' 2021.
\newblock \url{https://indico.fnal.gov/event/46752/contributions/224118/attachments/147660/189236/MCa__MUC_Targetry_Snowmass_23Sept2021.pdf}.

\bibitem{Alekou:2013eta}
A.~Alekou {\em et~al.} in {\em {Community Summer Study 2013}: {Snowmass on the Mississippi}}.
\newblock 10, 2013.
\newblock \href{http://arxiv.org/abs/1310.0804}{[{\ttfamily 1310.0804}]}.

\bibitem{Witte:2012zza}
H.~Witte {\em et~al.} Conf. Proc. C {\bfseries 1205201} (2012) 79--81.

\bibitem{Sato:2008zze}
A.~Sato {\em et~al.} Conf. Proc. C {\bfseries 0806233} (2008) THPP007.

\bibitem{hoeferkamp2022novelsensorsparticletracking}
M.~R. Hoeferkamp {\em et~al.}, 2022.
\newblock \url{https://arxiv.org/abs/2202.11828}.

\end{thebibliography}\endgroup

\end{document}